\documentclass[9pt,conference,a4paper,final]{IEEEtran}
\IEEEoverridecommandlockouts
\usepackage{multirow}
\usepackage{graphicx}
\usepackage{amsmath}
\usepackage{amssymb}
\usepackage{amsxtra}
\usepackage{threeparttable}

\usepackage[english]{babel}

\usepackage{amsmath}

\usepackage{amsthm}
\usepackage{amssymb}
\usepackage{amsfonts}
\usepackage{dsfont}
\usepackage{accents}
\usepackage{bm}

\usepackage{tabularx} 
\usepackage{longtable}
\usepackage{multirow}
\usepackage{booktabs}
\usepackage{makecell}
\usepackage{caption}
\usepackage{subcaption}
\usepackage{tcolorbox}

\usepackage{tikz,pgfplots}
\usepgfplotslibrary{groupplots}
\usetikzlibrary{intersections,calc,arrows,matrix,spy}
\usepackage{color}
\definecolor{darkred}{rgb}{0.6,0,0}
\definecolor{darkgreen}{rgb}{0,0.5,0}
\definecolor{darkblue}{rgb}{0,0,0.5}
\definecolor{SkyBlue}{rgb}{0.53, 0.81, 0.92}
\pgfplotsset{compat=1.5.1}

\usepackage{algorithm}
\usepackage{algpseudocode}
\makeatletter
\let\OldStatex\Statex
\renewcommand{\Statex}[1][3]{%
	\setlength\@tempdima{\algorithmicindent}%
	\OldStatex\hskip\dimexpr#1\@tempdima\relax}
\makeatother

\usepackage[squaren,Gray]{SIunits}

\def\R{\mathbb{R}}          									 	          

\def\e{\mathrm{e}}          									        	  

\newcommand{\pd}{\mathrm{\mathbf{p}}}

\newcommand{\xid}{\mathrm{\bm{\xi}}}
\newcommand{\rhod}{\mathrm{\bm{\rho}}}
\newcommand{\thetad}{\mathrm{\bm{\theta}}}

\newcommand{\Rd}{\mathrm{\mathbf{R}}}
\newcommand{\Ad}{\mathrm{\mathbf{A}}}

\newcommand{\Fd}{\mathrm{\mathbf{F}}}
\newcommand{\Id}{\mathrm{\mathbf{I}}}

\newcommand{\yd}{\mathrm{\mathbf{y}}}
\newcommand{\xd}{\mathrm{\mathbf{x}}}
\newcommand{\sd}{\mathrm{\mathbf{s}}}

\newcommand{\ud}{\mathrm{\mathbf{u}}}

\renewcommand{\Id}{\mathrm{Id}}

\newcommand{\Wd}{\mathrm{\mathbf{W}}}

\newcommand{\WP}{\mathrm{WP}}
\newcommand{\SNR}{\mathop{\mathrm{SNR}}}

\newcommand{\proj}{\mathbf{P}}
\newcommand{\rot}{\mathbf{R}}

\newcommand{\var}{\mathrm{Var}}

\def\NN{\mathbb{N}}

\def\SO{\mathrm{SO}}

\newcommand{\Lc}{\mathcal{L}}
\newcommand{\Gc}{\mathcal{G}}
\newcommand{\Sc}{\mathcal{S}}

\newcommand{\Uc}{\mathcal{U}}

\newtheorem{proposition}{Proposition}

\newcommand{\eqdef}{\ensuremath{\stackrel{\mbox{\upshape\tiny def.}}{=}}}

\definecolor{ao}{rgb}{0.0, 0.5, 0.0}
\usepackage{hyperref}

\newcommand{\revision}[1]{{\leavevmode\color{black}{#1}}}

\definecolor{mycolor1}{RGB}{47, 89, 245}%
\definecolor{mycolor2}{RGB}{214, 101, 41}%
\definecolor{mycolor3}{rgb}{0.,0.,1}%
\usepackage{float}

\graphicspath{{./Images/}}
\def\lw{0.5pt}

\usepackage{enumitem}

\setitemize{label=\textbullet, leftmargin=*, nolistsep}

\usepackage[left=1.62cm,right=1.62cm,top=1.9cm,bottom=4.45cm]{geometry}

\begin{document}


\makeatletter
\newcommand{\linebreakand}{%
  \end{@IEEEauthorhalign}
  \hfill\mbox{}\par
  \mbox{}\hfill\begin{@IEEEauthorhalign}
}
\makeatother

\title{Manifold Rewiring for Unlabeled Imaging
\thanks{This research was supported by the European Research Council (ERC) Starting Grant 852821$-$SWING.}
}

\author{
\IEEEauthorblockN{Valentin Debarnot}
\IEEEauthorblockA{\textit{DMI} \\
\textit{University of Basel}\\
Basel, Switzerland \\
valentin.debarnot@unibas.ch}
\and
\IEEEauthorblockN{Vinith Kishore}
\IEEEauthorblockA{\textit{DMI} \\
\textit{University of Basel}\\
Basel, Switzerland \\
vinith.kishore@unibas.ch}
\and
\IEEEauthorblockN{Cheng Shi}
\IEEEauthorblockA{\textit{DMI} \\
\textit{University of Basel}\\
Basel, Switzerland \\
cheng.shi@unibas.ch}
\and
\IEEEauthorblockN{Ivan Dokmani\'{c}}
\IEEEauthorblockA{\textit{DMI} \\
\textit{University of Basel / UIUC}\\
Basel, Switzerland / Urbana, IL, USA \\
ivan.dokmanic@unibas.ch}
}

\maketitle
\thispagestyle{empty}

\begin{abstract}
Geometric data analysis relies on graphs that are either given as input or inferred from data. These graphs are often treated as ``correct'' when solving downstream tasks such as graph signal denoising. But real-world graphs are known to contain missing and spurious links. Similarly, graphs inferred from noisy data will be perturbed. We thus define and study the problem of graph denoising, as opposed to graph signal denoising, and propose an approach based on link-prediction graph neural networks. We focus in particular on neighborhood graphs over point clouds sampled from low-dimensional manifolds, such as those arising in imaging inverse problems and exploratory data analysis. 
We illustrate our graph denoising framework on regular synthetic graphs and then apply it 
to single-particle cryo-EM where the measurements are corrupted by very high levels of noise. Due to this degradation, the initial graph is contaminated by noise, leading to missing or spurious edges. We show that our proposed graph denoising algorithm improves the state-of-the-art performance of multi-frequency vector diffusion maps.

\end{abstract}

\section{Introduction}

The manifold assumption refers to high dimensional datasets that can be described by a small number of parameters living on a manifold. Manifold structure arises in diverse applications, from spaces of optical blurs in microscopy \cite{debarnot2020learning}, over patches of natural images \cite{carlsson2008local}, to the behavior of neuronal populations in the hippocampus \cite{gardner2022toroidal}.

Once the manifold assumption is established, the natural question is that of manifold learning: how to characterize a manifold from a noisy set of points sampled from it \cite{belkin2008towards, balasubramanian2002isomap}.
Most approaches begin by constructing a neighborhood graph that links data points that are close in some well chosen metric. These observed data points are often corrupted by noise which introduces spurious or missing edges in the graph and hampers the estimation of the manifold. 

In this paper, we address the problem of denoising neighborhood graphs sampled from manifolds with a focus on imaging problems where the observations suffer from a large amount of noise. The denoising takes the form of a rewiring of the noisy graph obtained from the observations. We build on the recent advances in graph neural networks for link prediction. An important example of an imaging modality where the observations lie on a manifold 
is single-particle cryo-electron microscopy (cryo-EM, 2017 Nobel prize in Chemistry)
which obtains atomic-resolution images of biological proteins. A major challenge in cryo-EM is noise. As we show later in Section \ref{sec:experiments}, our plug-in graph denoising algorithm can improve performance of state-of-the-art methods.

\section{Related work}

\subsection{Cryo-EM reconstruction}

Approaches to unknown-view tomographic reconstruction in cryo-EM can be categorized according to whether or not they estimate the orientations of the projections. Those that don't estimate the projections typically employ the method of moments  \cite{kam1980reconstruction,levin20183d,sharon2020method,huang2022orthogonal}. Our focus is, however, on the methods that estimate orientations which achieve the current state-of-the-art performance at low signal to noise ratios (SNR) \cite{punjani2017cryosparc,zhong2021cryodrgn2}. Such methods, which typically alternate between orientation estimation and reconstruction steps, are at the heart of major software packages for cryo-EM reconstruction such as RELION \cite{scheres2012relion}.
A major challenge faced by these approaches is that huge noise hampers orientation estimation and precludes reconstruction at very low SNRs if only pairwise information between observations is used. Nevertheless, leveraging multiple projections and sophisticated denoising schemes enables sufficiently accurate angle estimation. In this paper, we build on this line of work to increase the maximum tolerable amount of noise.

\subsection{Graphs and tomography}

The problem of 2D computed tomography with projections at unknown angles shares many similarities with single-particle cryo-EM.
In the seminal paper \cite{coifman2008graph}, the authors show that it is possible to retrieve these unknown orientations from the noisy observations. This is done by constructing a graph with projections as vertices and edges between those projections that are close in the $\ell_2$ metric.
The first two non-trivial eigenvectors of the corresponding graph Laplacian encode relative position of the projections.
This feat can be explained by linking the eigenfunctions of the Laplace--Beltrami operator on the circle to the eigenvectors of the graph Laplacian. The authors of \cite{singer2013two} proposed to augment projection denoising with graph denoising based on the Jaccard index. This combination allows reconstruction at very high noise levels. In this paper we look at more general graph denoising strategy based on graph neural networks and the considerably more complex full 3D case.


\revision{The current state-of-the-art methods for reconstruction at very low SNR combine signal filtering with the construction of a graph as in 2D computed tomography} \cite{fan2019multi,zehni2020joint,fan2021cryo}. 
\revision{Each projection is filtered by using a low-dimensional representation basis} \cite{zhao2016fast,zhao2013fourier,landa2017steerable} and denoised by averaging with projections at similar viewing directions  \cite{zhao2014rotationally,singer2011viewing}.
Similarly to the 2D case, the success of the averaging step relies on the quality of the neighborhood graph, which is in turn impacted by the large amount of noise in projection images. To improve performance, the state-of-the-art methods additionally leverage vector diffusion maps (VDM) to penalize inconsistencies among sequences of estimated transformations between projections \cite{singer2012vector}. 

The VDM embedding can be further enriched by considering multiple irreducible representations of the 2D rotation group, resulting in the so-called multi-frequency vector diffusion maps (MFVDM) \cite{fan2019multi,zehni2020joint,fan2021cryo}.
VDM and MFVDM compute embeddings that are robust even in the presence of high noise.

\subsection{Denoising}

Naive cryo-EM denoising methods rely on unsupervised classification approaches such as K-means \cite{frank2006three}, and averaging in-class projections. However, accounting for the symmetries requires the costly alignment of all 2D projections.
The authors of \cite{zhao2013fourier} use the Fourier-Bessel basis to efficiently apply rotations and flips. The Fourier-Bessel basis was also used to improve classification and averaging  \cite{zhao2014rotationally} as it allows a fast computation of the \revision{VDM embedding}. This improved nearest neighbor search is then used to denoise projections by averaging their Fourier--Bessel expansion coefficients   \cite{zhao2014rotationally}.
Recent work leverages the MFVDM embedding and proposes refined denoising schemes that use the eigenvalues and eigenvectors of the MFVDM matrix to filter the Fourier-Bessel coefficients \cite{fan2019multi,fan2021cryo}.

\subsection{Contributions}

In this paper, we propose a new method to denoise neighborhood graphs affected by noise. Our method is plug-and-play and can be combined with any existing graph-based scheme like VDM or MFVDM. We build on a state-of-the-art GNN-based graph link prediction algorithm \cite{pan2021neural} to remove edges that are inconsistent with the learned prior graph statistics. 
We show that the proposed algorithm indeed produces a better graph and thus improves denoising and reconstruction.
Combined with the MFVDM-based reconstruction pipeline, it  achieves the state-of-the-art in low-SNR cryo-EM on synthetic data generated from real densities. 

\section{Methods}
We use bold typeface for vectors $\xd$ and matrices $\Ad$, and regular typeface for scalars and scalar functions.
The $i$-th entry of a vector $\xd$ is written $x_i$ or $\xd[i]$.  $\Ad^\intercal$ is the transpose of $\Ad$.
We let $\Uc(I)$ denote the uniform distribution over the set $I$.
The 2-sphere $\Sc^2$ is defined as
\begin{equation*}
    \Sc^2 \eqdef \{\xd\in\R^3; \|\xd\|_2 = 1 \}.
\end{equation*}
We let $\SO(d)$ denote the rotation group in $d$ dimensions
Elements of $\SO(d)$ can be identified with $d \times d$ orthogonal matrices with determinant $+1$,
\begin{equation*}
    \SO(d) = \{ \Rd \in\R^{d\times d} \ ; \ \det(\Rd)=1,\  \Rd^\intercal\Rd = \Id_d \},
\end{equation*}
where $\Id_d\in\R^{d\times d}$ is the identity matrix. For simplicity we write $\Rd(\thetad)$ for both the $d \times d$ rotation matrix with angles $\thetad\in\R^d$ acting on vectors in $\mathbb{R}^d$ and the rotation operator acting on the discrete $d$-dimensional volume $\R^{n^d} = \R^{n  \times \cdots \times n}$.\footnote{Foregoing technicalities, this operator is then the left regular representation of the rotation group on $\ell^2(\mathbb{Z}_n^d)$.}

\subsection{Single-particle cryo-EM}
Single-particle cryo-EM aims to recover volume density of a single  molecule at atomic resolution.
To this end, a solution with many instances of the molecule to be imaged is placed on a grid and rapidly frozen. Then an electron beam passes through the randomly oriented molecules and the resulting two-dimensional projections along the beam are captured by a camera. Finally, single-particle projections are extracted using automated tools \cite{frank1983automatic,heimowitz2018apple}. This is equivalent to observing projections of the 3D density along random directions followed by a rotation around the projection axis; see Fig. \ref{fig:cryoEM_pipeline}. 

\begin{figure}[h]
    \begin{center}
    \includegraphics[width=8cm]{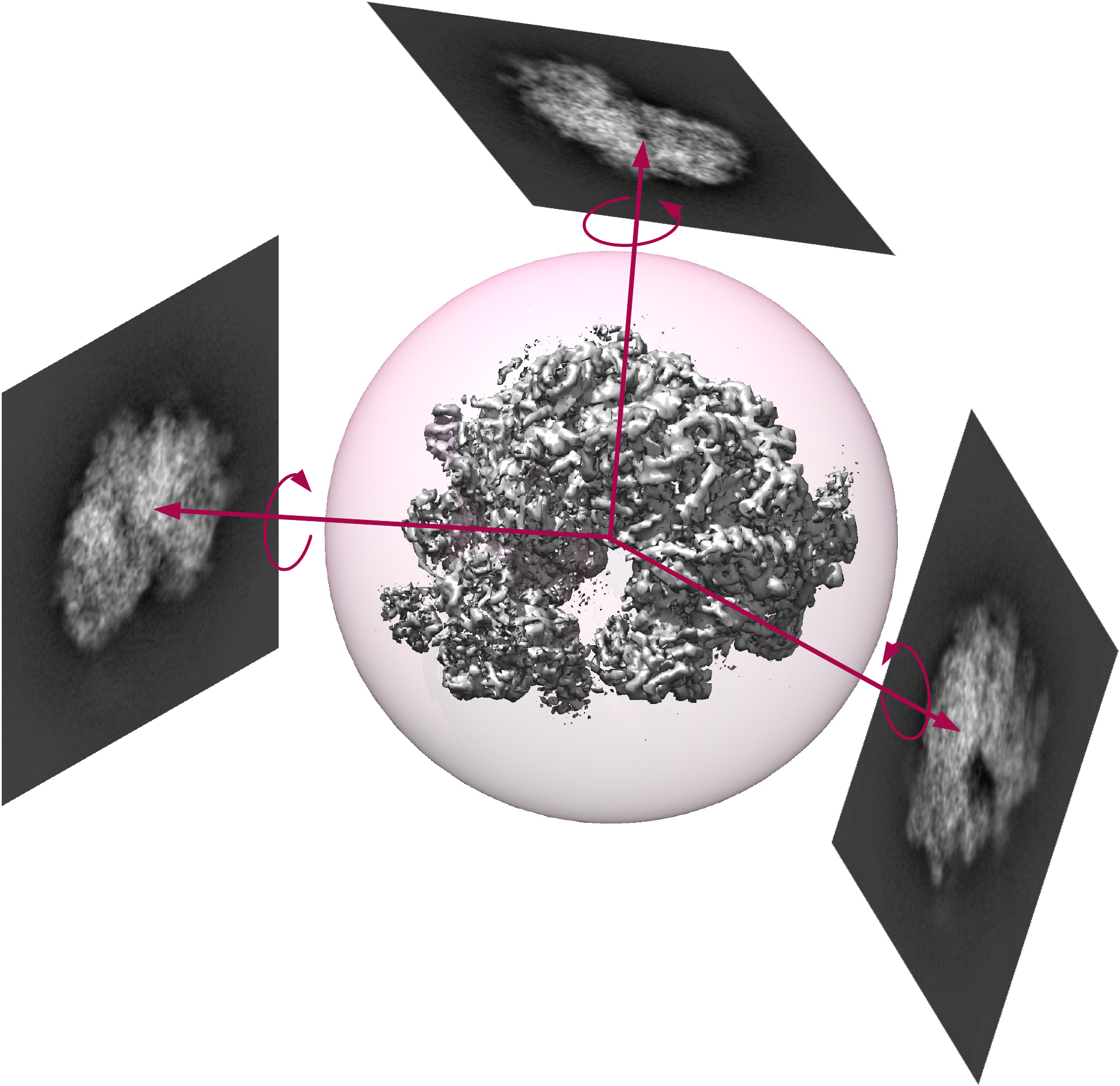}
    \end{center}
	\caption{Illustration of single-particle cryo-EM imaging. We observe projections of the volume density along random directions. For a fixed viewing direction, any rotation of the molecule around this axis will give rise to the same projection up to a 2D rotation.  \label{fig:cryoEM_pipeline}} 
\end{figure}

Mathematically, the cryo-EM reconstruction problem boils down to estimating an unknown volume $\rhod\in\R^{n\times n \times n}$ with $n\in\NN_+$, from $M\in\NN_+$ projections at unknown random angles, corrupted by noise. Let $1\leq m \leq M$. The pixel intensity $\yd_m\in\R^{n\times n}$ of the $m$-th observed projection image is given by
\begin{equation}
    \label{eq:problem_definition}
    \yd_m = \proj \rot(\thetad_m) \rhod + \xid_m,
\end{equation}
where we use $ \rot(\thetad_m)\in\SO(3)$ to denote the Euler rotation angles that parameterize the rotation $\Rd(\thetad_m)$, $\proj$ is a projection operator from $\R^{n\times n\times n}$ to $\R^{n\times n}$ which sums over the third ($z$) coordinate, and $\xid_m$ denotes independent additive white Gaussian noise with covariance  $\sigma^2\Id_{n^2}$.
In single-particle cryo-EM, it is reasonable to assume that molecules have no preferred orientation. We therefore assume that the particle rotations are sampled uniformly and independently at random over $\SO(3)$.

We observe a large number of projections, typically $M\geq 10,000$. However, because the electron dose must be kept low in order not to destroy the molecules, observations suffer from a lot of noise. 
The noise degradation between a clean signal $\yd_0\in\R^n$ and a noisy signal $\yd_0+\xid\in\R^n$ is measured by the SNR,
\begin{equation*}
    \SNR(\xid,\yd_0) = \frac{\var(\yd_0)}{\var(\xid)}.
\end{equation*}
In cryo-EM, it is not unusual to encounter SNRs as low as 0.01; We are particularly interested in this high noise regime.
Even at such low SNRs, the volume $\rhod$ in Problem \eqref{eq:problem_definition} can be accurately estimated if the orientations $(\thetad_m)$ are known. If that is the case the problem becomes a linear tomographic inversion  and can be solved by the pseudoinverse of the forward map, that is the filtered backprojection.
The challenge is that since the molecules are arranged randomly, we do not \textit{a priori} know the orientations.

\subsection{Affinity graph}

State-of-the-art projection denoising methods construct a graph where each vertex correspond to one noisy observation \cite{singer2012vector,fan2021cryo}. The starting point to construct the graph is an affinity matrix $\Wd\in\R^{M\times M}$ between projections
\begin{equation*}
    \Wd_{m,m'} = \mathrm{affinity}(\yd_m,\yd_{m'}),
\end{equation*}
where the $\mathrm{affinity}$ function varies from method to method.

Early work uses the  rotation-invariant $\ell^2$ metric and then refines the affinity using the connection graph Laplacian \cite{singer2012vector}. In the state-of-the-art MFVDM the affinity is based on the eigenvectors and eigenvalues of connection graph Laplacian corresponding to multiple irreducible representations of the 2D rotation group. 
Both approaches construct the adjacency matrix of the graph by connecting a node to its $K$ nearest neighbors. We let $\Gc=(V,E)$ denote this graph and $\Ad\in\R^{M\times M}$ its adjacency matrix, where $V\subset\R^{n\times n}$ is the set of vertices and $E\eqdef \left\{ (i,j) \text{ if }i \text{ is connected to } j;  1\leq i,j \leq M  \right\}$ is the set of edges.

An important feature of the two discussed affinity functions is that they are invariant to the rotation of the molecule around the projection axis. 
Each observation being a function of a rotation of the molecule which lives in $\mathrm{SO}(3)$, this invariance amounts to quotienting the space of observations $\mathrm{SO}(3)$ by the space of 2D rotations $\SO(2)$, $\SO(3)\backslash \SO(2) \simeq \Sc^2$, \cite[Chapter 3, Problem 6]{rubakov2009classical}.
The affinity thus depends only on the difference between the projection axes (and on the density, of course) which are defined by unit vectors in $\R^3$, that is to say, vectors on $\mathcal{S}^2$. For sufficiently regular $\rhod$ without symmetries, this  implies that in the absence of noise the affinity graph will topologically resemble a 2-sphere. This is formalized by the following proposition proved in Appendix \ref{app:proof_prop}.

\begin{proposition}\label{prop:sphere}
Let $\Rd(\thetad) \sim \Uc(\SO(3))$ with  $\thetad=[\theta_{1},\theta_{2},\theta_{3}]$ being the Euler angles. Let further 
\begin{equation*}
    \yd = \proj \rot(\thetad) \rhod.
\end{equation*}
Then there exists a function $f_\rhod:\Sc^2\to \R^{n\times n}$ which depends on the unknown density $\rhod$, such that for $\phi\sim\Uc(\mathrm{SO}(2))$, $\sd\sim\Uc(\Sc^2)$,
\begin{equation}\label{eq:fct_sphere}
    \tilde{\yd} = \rot(\phi) f_\rhod(\sd)
\end{equation}
has the same distribution as $\yd$.
\end{proposition}

State-of-the-art affinity metrics further leverage a reflection identity: 
projections corresponding to opposite viewing directions coincide up to a flip. 
This fact must be accounted for when we later design the graph denoising strategy.

Topologically, it means that observations should live on a projective plane rather than the sphere after quotienting out the flips. This is however only true for noiseless observation. As the noise level increases, the constructed graph is contaminated by a growing number of spurious and missing edges, as illustrated in Fig. \ref{fig:graph_noisy}.
In the next section, we describe Walk Pooling \cite{pan2021neural}, a link prediction graph neural network, and demonstrate how it can be used to improve the noise resilience of existing approaches.

\begin{figure*}[t]
    \centering
	\begin{tikzpicture}
	\node at (0,0) { \includegraphics[scale=0.28]{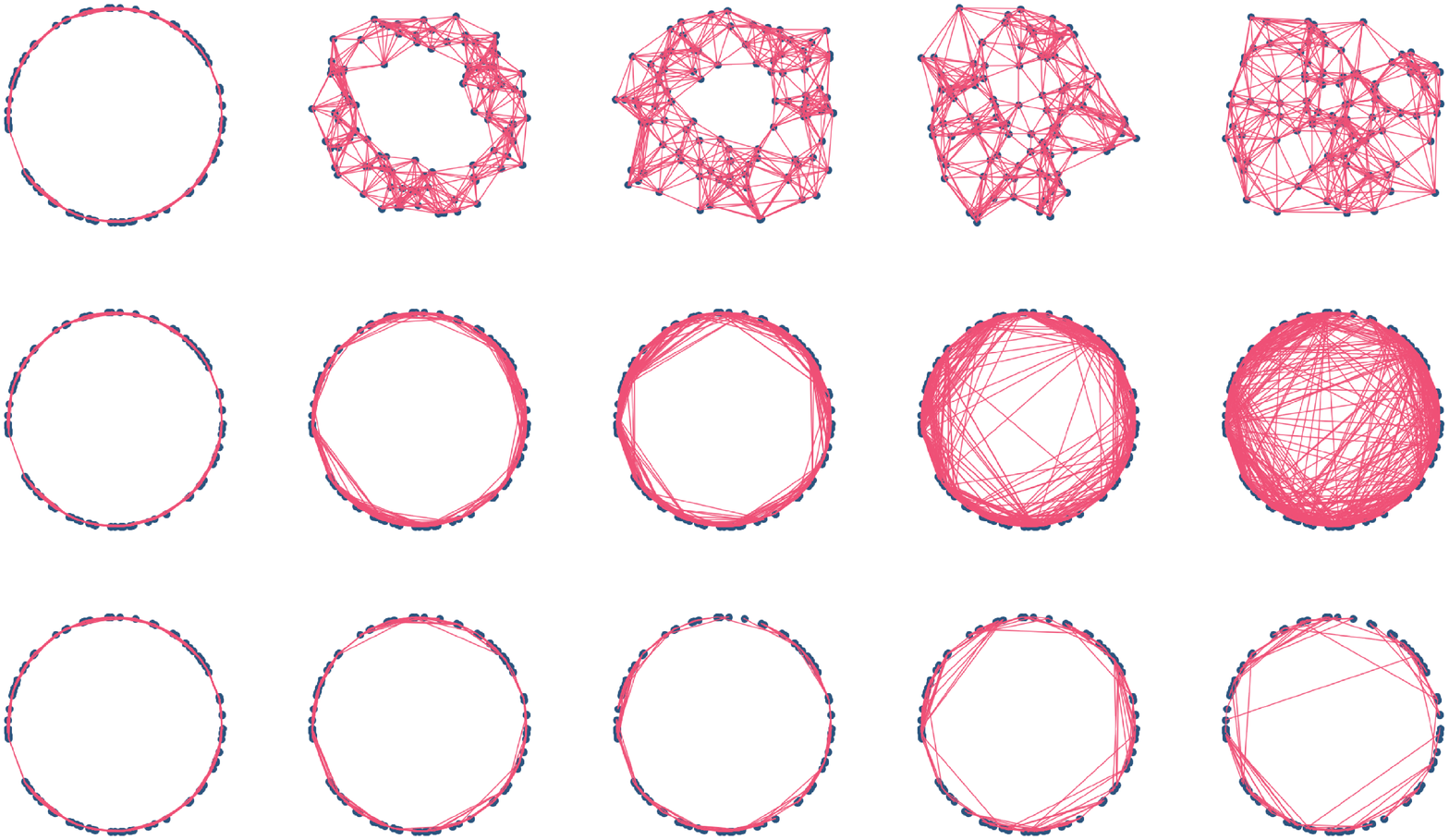}  };
	\node at (8.5,0) { \includegraphics[scale=0.35]{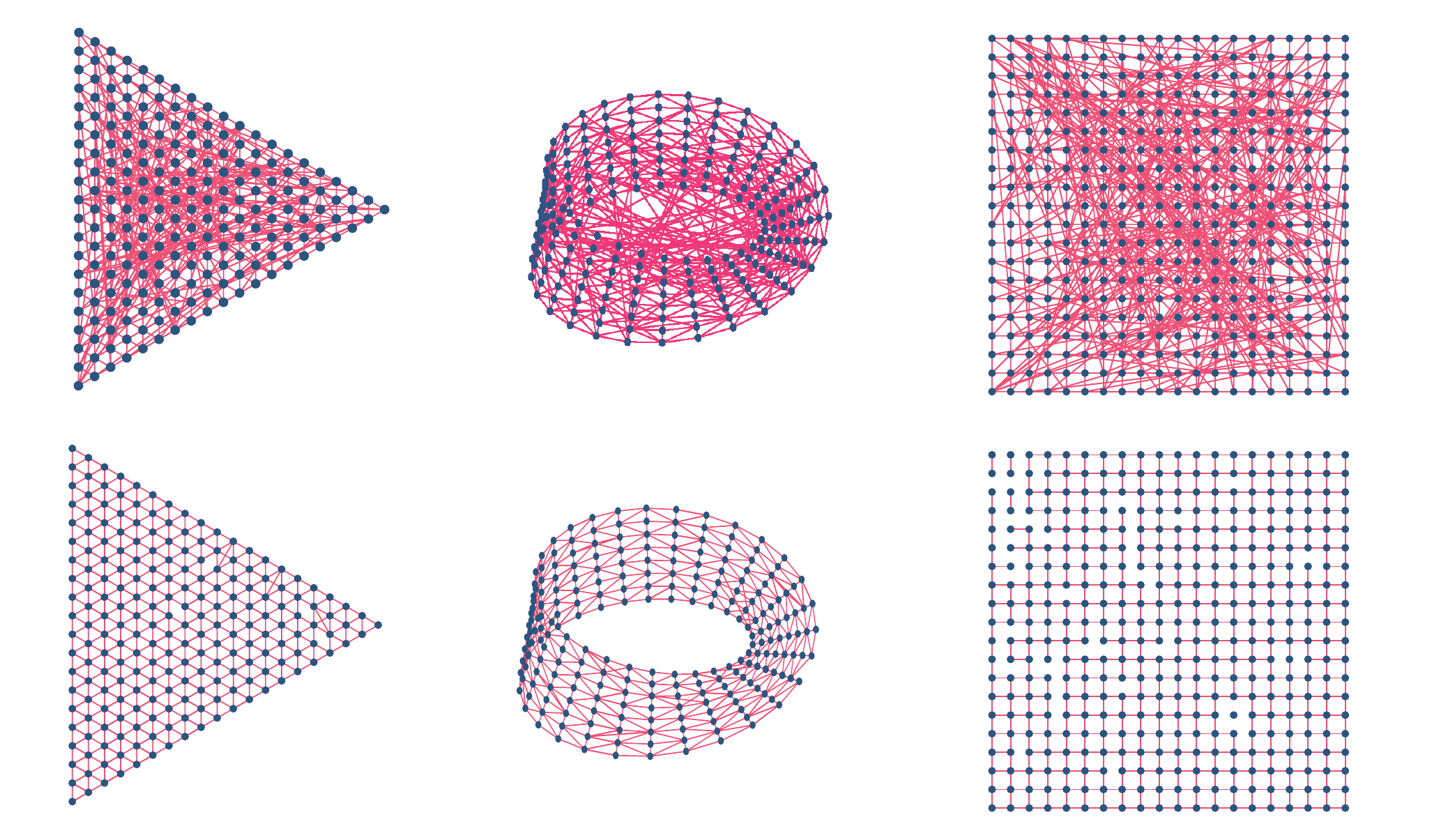}  };
	\end{tikzpicture}
	\caption{
	Illustration of WP-based graph denoising on a graph sampled from a circle manifold and regular grid graphs. 
	\textbf{Left}: We  randomly sample 100 points on the circle (top-left) and add  noise to their positions with increasing amplitude (first row). The noisy graph is constructed from 10 nearest neighbors.
	The second row visualizes these graphs with nodes in the original positions. As in cryo-EM, with large perturbations, faraway nodes become connected and destroy the graph geometry. The goal of graph denosing is to remove these edges based on the prior knowledge that the nodes of true graph are randomly sampled from a circle. The third row shows the denoised graphs.
	\textbf{Right}: Nodes (blue) are shown at their true positions, but the graph contains many spurious links due to noise. The graphs are sampled on (from left to right) a triangle grid on the 2D plane, a triangle grid on the M\"obius strip and a square grid on the 2D plane.
	The second row shows the denoised graph after WP. We see that WP yields meaningful denoised graphs even in the presence of large noise.
	\label{fig:graph_noisy}} 
\end{figure*}

\subsection{Walk pooling: likelihood of the correct graph (subgraph)}

In this section we introduce Walk Pooling (WP) \cite{pan2021neural}, a graph neural network that is trained to output the likelihood that an edge belongs to a target graph. Fig. \ref{fig:graph_noisy} illustrates the results of applying WP denoising to synthetic regular graphs.

Following the discussion in the previous section and Proposition \ref{prop:sphere}, the graph $\Gc$ is expected to have the same distribution as a $K$-neighborhood graph sampled from the sphere $\Sc^2$ (or projective plane if the affinity metric identifes the flips). We let $\bar{\mathcal{G}}=(\bar V, \bar E)$ denote this target graph. 
To construct $\bar\Gc$, we randomly sample $M$ points from the unit 2-sphere and compute the $K$-neighbourhood graph using the geodesic distance. 

Let $h\in\NN_+$. We define the $h$-hop sub-graph of $\Gc$ corresponding to an edge $e\in E$ by $\mathcal{G}^h_{e} \eqdef (V^h_e,E^h_{e})$ where
\begin{equation*}
    E^h_{e} \eqdef \{\e' \in E, \text{ there is at most }h-1\text{ edges between }e' \text{ and } e \},
\end{equation*}
and $V_e^h\subset V$ the set of vertices that have an edge in $E_e^h$.

The distribution of sub-graphs depends on the manifold that the data points are sampled from. Since $\Sc^2$ is isotropic with constant sectional curvature, we expect that the sub-graph distribution is invariant to the position in the graph. While the sub-graph distribution for regular random graphs (such as regular neighborhood graphs on the sphere) could at least in principle be characterized, explicitly calculating posterior probabilities of link existence is impossible.
This difficulty suggests the use of machine learning algorithms such as WP to approximate the optimal estimate of the link or sub-graph likelihood.

More concretely, we want to learn a score function 
\begin{equation*}
    s(e| \mathcal{G})=\WP_\pd(\mathcal{G}^h_{e} ),
\end{equation*}
where $\WP_\pd:(e\in E,\Gc) \to [0,1]$ is a neural network parameterized by $P$ trainable parameters $\pd\in\R^{P}$. 
The WP formalism computes permutation invariant features from a sub-graph from powers of its adjacency matrix \cite{pan2021neural}.
It then applies a multilayer perceptron (MLP), such that 
\begin{equation*}
    \WP_\pd(\mathcal{G}^h_{e} ) = \mathrm{MLP}_\pd(\Fd_{e}),
\end{equation*}
where $\Fd_{e}\in \mathbb{R}^{2\times (2h-1)}$ are the mentioned input features. The matrix $F_e$ is defined as 
\begin{equation}\label{eq:fetaure}
    \Fd_{e} \eqdef \left[
    \arraycolsep=1.4pt\def\arraystretch{1.3}\begin{array}{cccc}
     \left({\Ad^+}_{{\mathcal{G}_{e}^h} }\right)^2_{01}& \left({\Ad^+}_{{\mathcal{G}_{e}^h} }\right)^3_{01}& \cdots&\left({\Ad^+}_{ {\mathcal{G}_{e}^h}}\right)^{2h}_{01}\\
    \vspace{0.2cm}
    \left({\Ad^-}_{ {\mathcal{G}_{e}^h}}\right)^2_{01}&\left({\Ad^-}_{ {\mathcal{G}_{e}^h}}\right)^3_{01}& \cdots&\left({\Ad^-}_{ {\mathcal{G}_{e}^h}}\right)^{2h}_{01}
    \end{array}
    \right],
\end{equation}
where 
\begin{itemize}
    \item $\Ad^+_{\mathcal{G}^h_{e}}$ is the sub-graph adjacency matrix of  $\mathcal{G}^{h,+}_{e} = (V,E^h_{e} \cup \{e\})$,
    \item $\Ad^-_{\mathcal{G}^h_{e}}$ is subgraph adjacency matrix of $\mathcal{G}^{h,-}_{e} = (V,E^h_{e} \backslash \{e\})$,
    \item $\left(\Ad\right)^k_{ij}$ is the element in $i$-row and $j$-column of $\Ad^k$.
    \item We assume that the vertices of the link of interest are in positions 0 and 1 in matrices $\Ad^\pm_{\mathcal{G}^h_{e}}$.
\end{itemize}

The MLP is trained using the target graph $\bar{\mathcal{G}}$. For all links in $\bar{\mathcal{G}}$, the WP neural network is trained to output a score of $1$. Then, we randomly sample the same number of negative links (not existing links) from $\bar{\mathcal{G}}$ and give them score $0$. This can be regarded as a binary classifier with training dataset $\{\Fd_{e},s\} $. After training this MLP, we can then compute the score for the links in $\bar{\mathcal{G}}$.

\subsection{Cryo-EM denoising with Walk Pooling}

The spurious noisy edges in the neighborhood graph deteriorate the quality of projections denoised by averaging over neighbors. We use WP to detect and remove these erroneous faraway connections from the affinity graph $\Gc$. We emphasize that WP can be combined with \textit{any} existing graph-based method (e.g., VDM or MFVDM).

Let $\tau>0$ be the threshold. For every edge $e\in E$ in the noisy graph $\Gc$, we compute the score $ s(e| \mathcal{G})$. We then construct the denoised graph $\hat \Gc = (\hat V, \hat E)$ by keeping all edges with a score above  $\tau$. 
If all edges of a vertex have been removed, we removed this vertex from the graph, such that $\hat V \subset V$ and $\hat E\subset E$.

\section{Computer experiments}
\label{sec:experiments}

We now demonstrate that our approach improves the  denoising performance and consequently of reconstruction of state-of-the-art methods. 
Concretely, we use the ASPIRE toolbox \cite{zhao2014rotationally} and MFVDM denoising \cite{fan2021cryo} with and without our graph denoising step. 
Numerical comparisons in the original MFVDM publication \cite{fan2021cryo} show that below a SNR of 0.05, MFVDM performs better than standard methods in single particle cryo-EM \cite{scheres2012relion,zhao2014rotationally,bhamre2016denoising,landa2017steerable}. Therefore, we only compare our graph denoising pipeline with this approach.

\subsection{Experimental setting}
We closely follow the experimental setting described in \cite{fan2021cryo}. We simulate cryo-EM by using a 3D density of a 70S ribosome discretized on a grid of size $129\times 129 \times 129$ ($n=129$).
We generate $M=10,000$ projections by sampling uniformly at random $M$ rotations angles $(\thetad_m)$ and add white Gaussian noise with of various power, such that the average SNR of the projections is $0.03,0.016,0.01$. (These values correspond to  $-15$dB, $-18$dB and $-20$dB.) The MFVDM approach constructs a $K$-NN graph on $K=50$ neighbors.

\def\wid{3.2cm}
\def\hs{-1cm}
\def\xx{3.5}
\def\yy{-4.5}
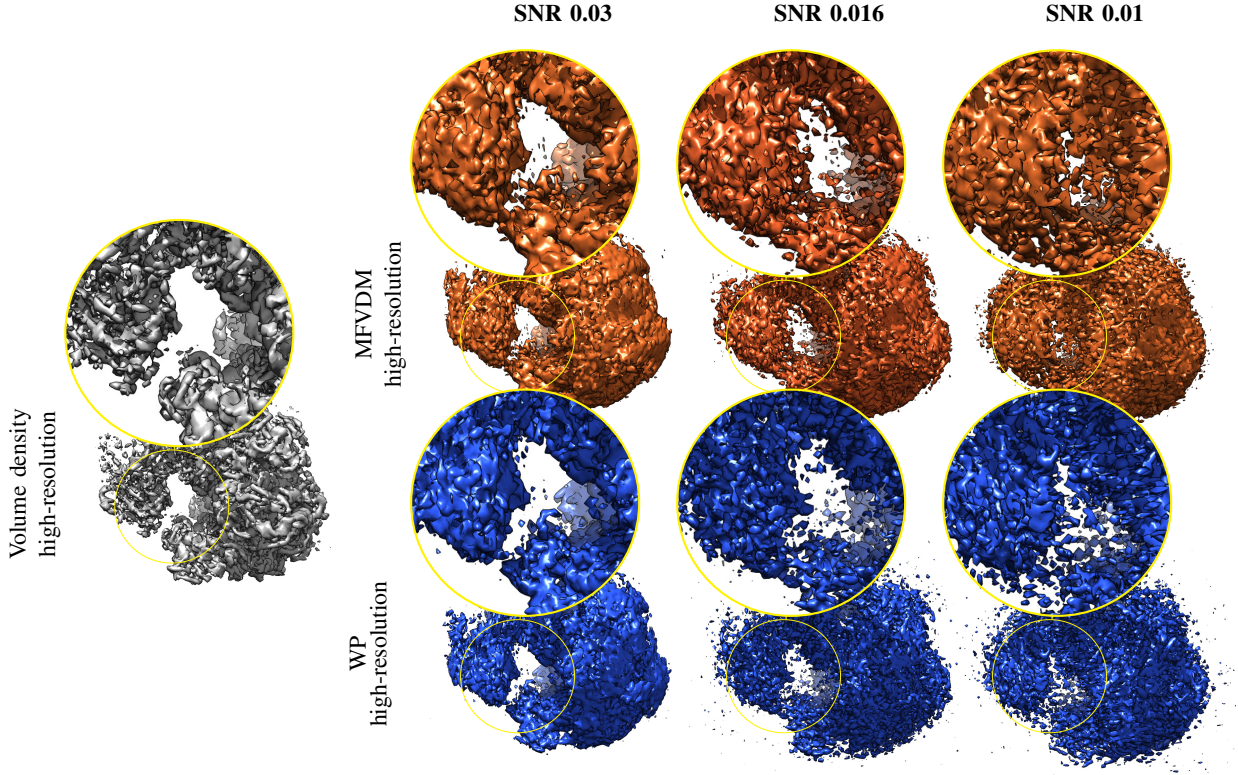
\begin{figure*}[h]
        \hspace{0cm}
        \centering
		\begin{tikzpicture}[spy using outlines={circle,yellow,magnification=2,size=3cm, connect spies}]
		\node[text width=2cm,text centered,minimum width=2cm,minimum height=2cm, rotate=90] at (-1.3*\xx-2.4,0.5*\yy) {Volume density high-resolution};
		\node[rotate=270] at (-1.3*\xx,0.5*\yy) { \includegraphics[width=\wid]{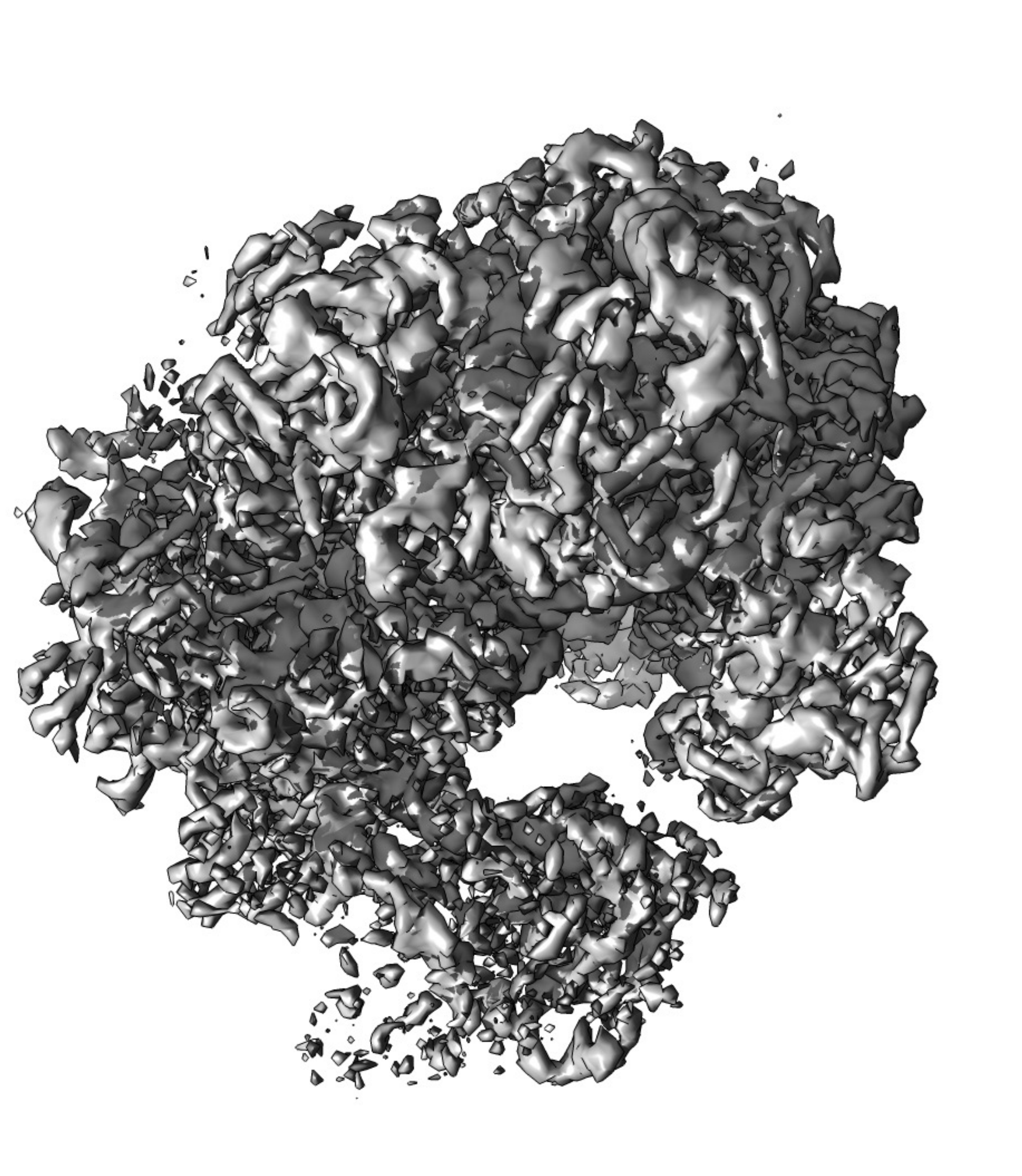}  };
		\node[text width=2cm,text centered,minimum width=2cm,minimum height=2cm] at (0.*\xx,4) {\textbf{SNR 0.03}};
		\node[text width=2cm,text centered,minimum width=2cm,minimum height=2cm] at (1*\xx,4) {\textbf{SNR 0.016}};
		\node[text width=2cm,text centered,minimum width=2cm,minimum height=2cm] at (2*\xx,4) {\textbf{SNR 0.01}};
		
		\node[text width=2cm,text centered,minimum width=2cm,minimum height=2cm, rotate=90] at (-2.4,0) {MFVDM high-resolution};
		\node[rotate=270] at (0*\xx,0) { \includegraphics[width=\wid]{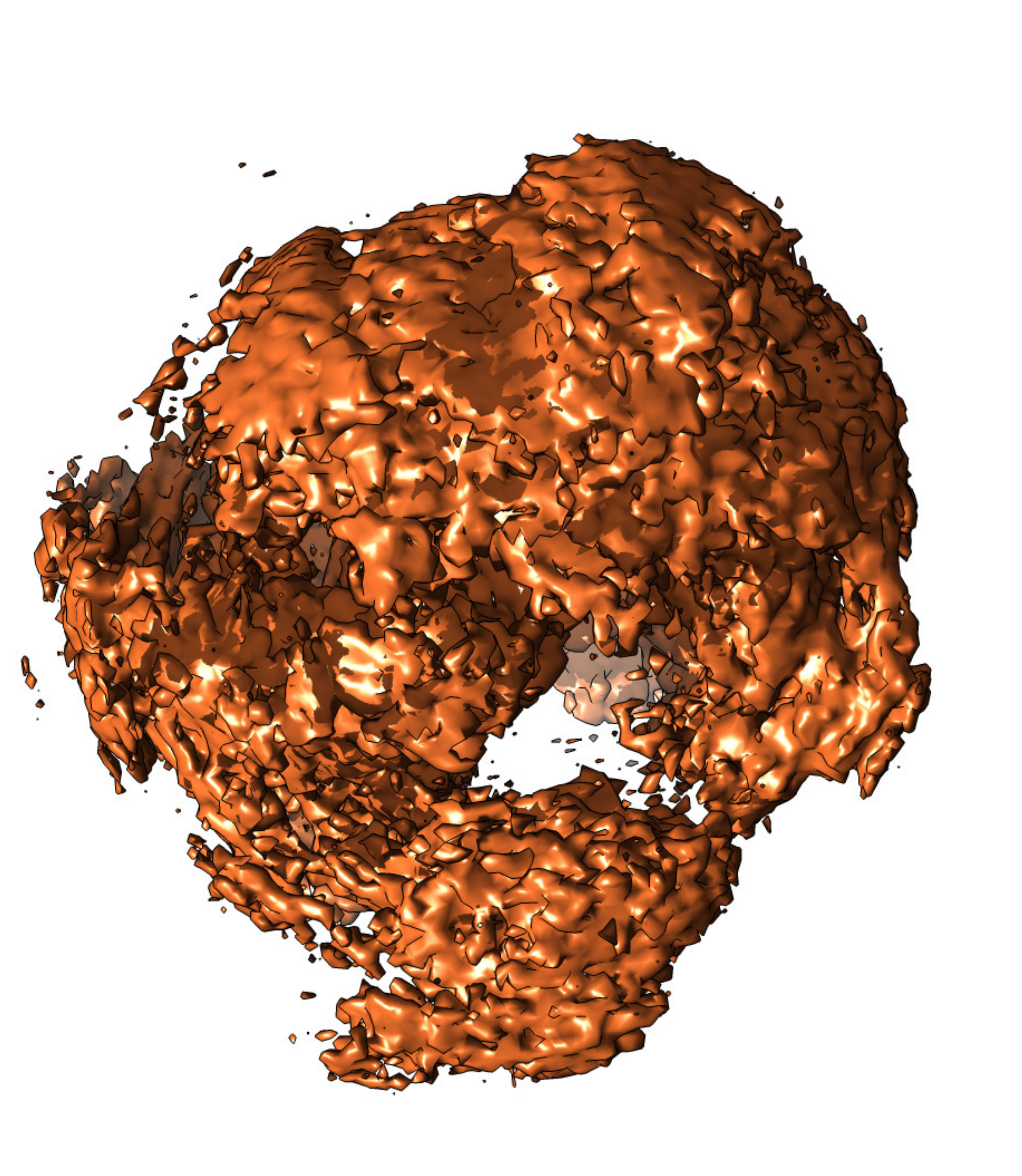}  };
		\node[rotate=270] at (1*\xx,0) { \includegraphics[width=\wid]{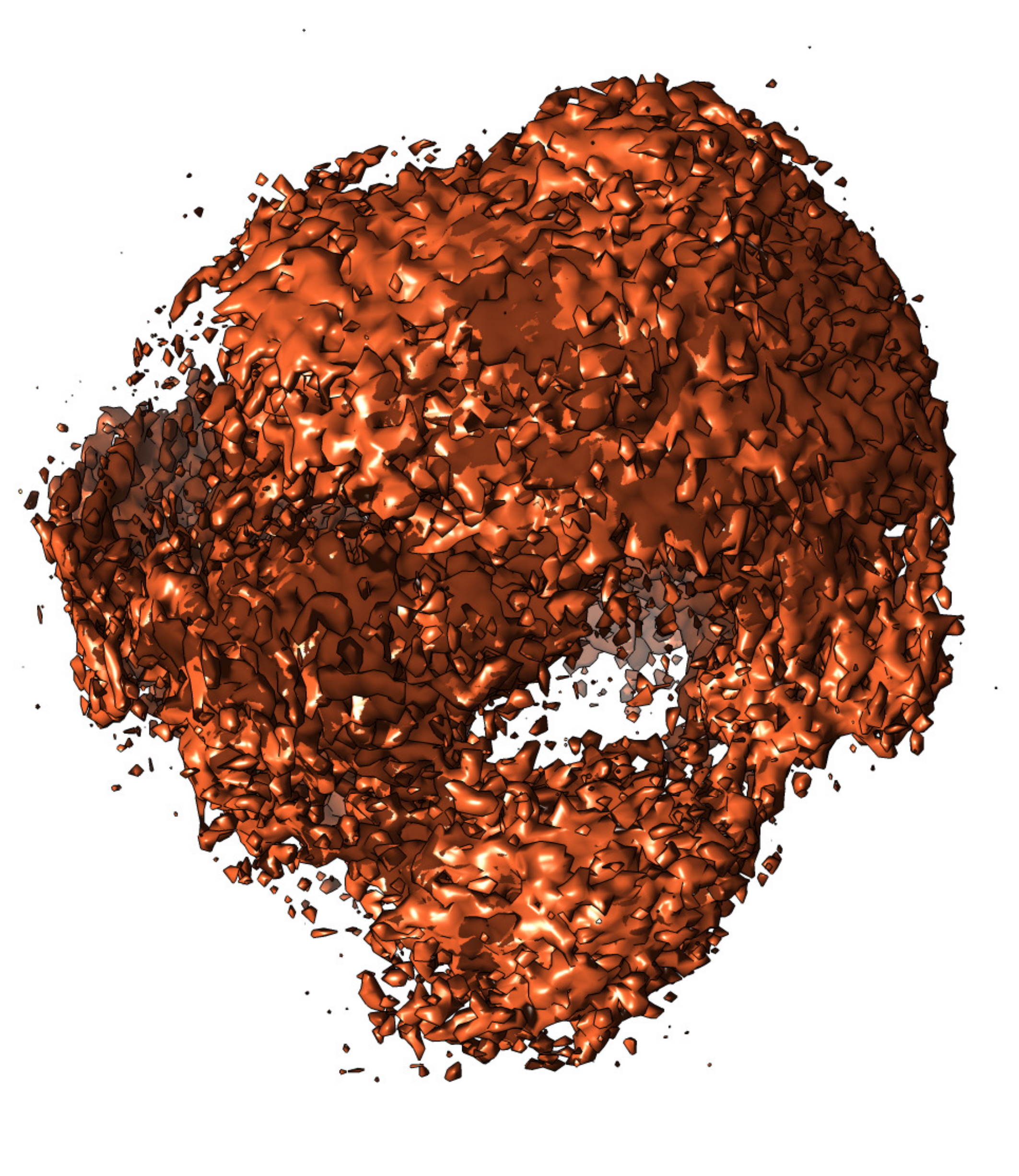}  };
		\node[rotate=270] at (2*\xx,0) { \includegraphics[width=\wid]{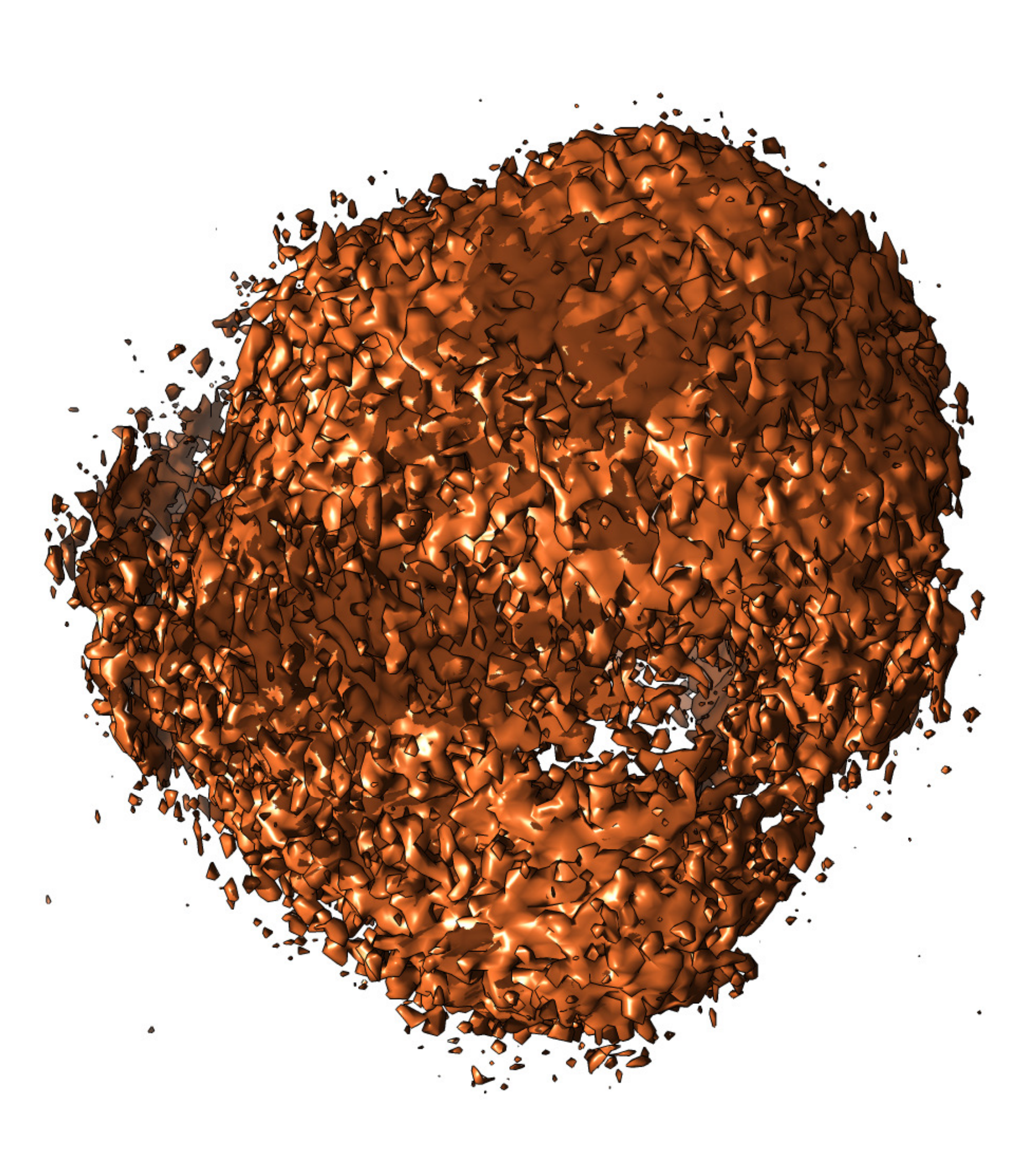}  };
		
		\node[text width=2cm,text centered,minimum width=2cm,minimum height=2cm, rotate=90] at (-2.7,\yy) {WP};
		\node[text width=2cm,text centered,minimum width=2cm,minimum height=2cm, rotate=90] at (-2.4,\yy) {high-resolution};
		\node[rotate=270] at (0*\xx+0.,\yy) { \includegraphics[width=\wid]{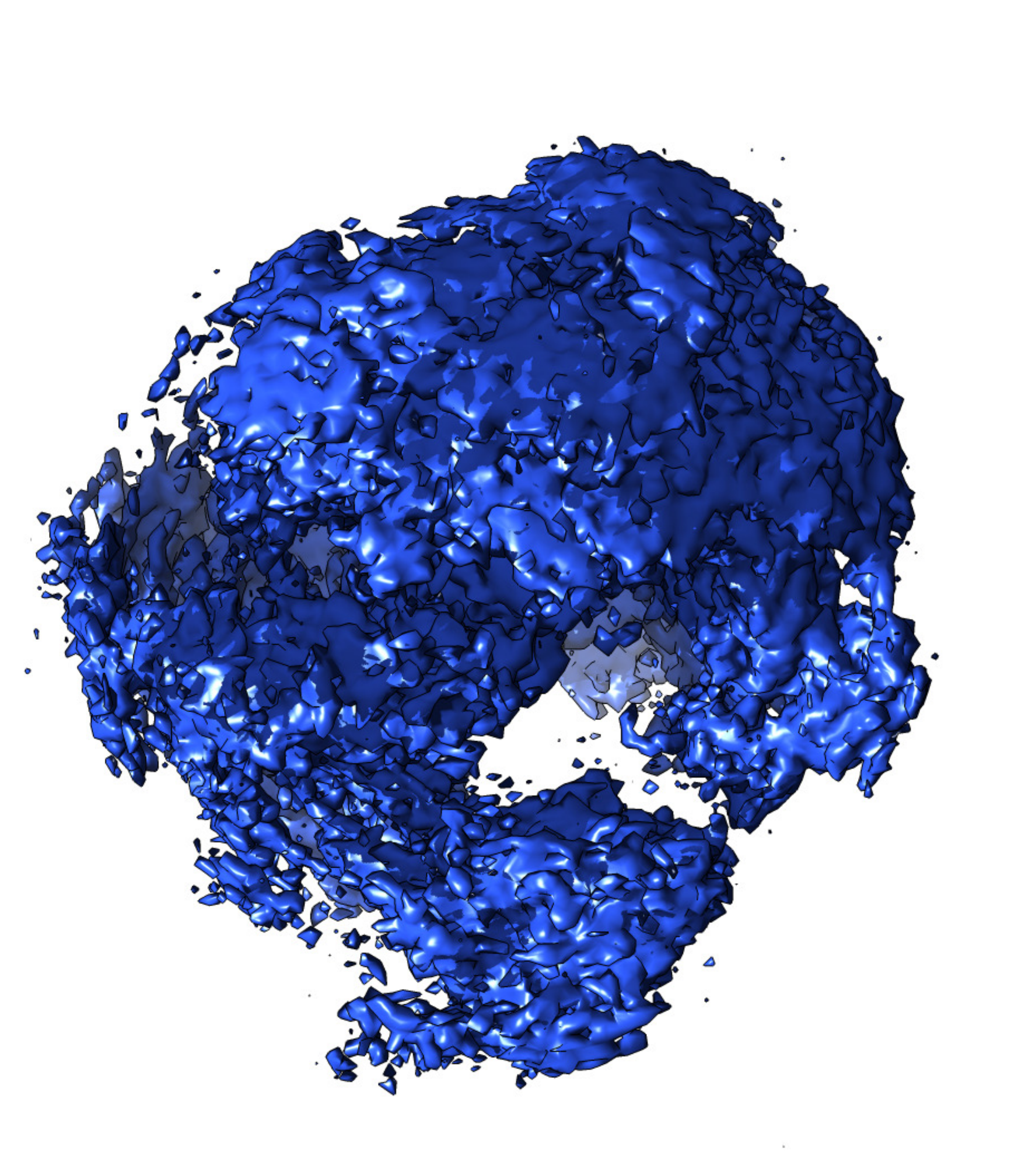}  };
		\node[rotate=270] at (1*\xx+0.,\yy) { \includegraphics[width=\wid]{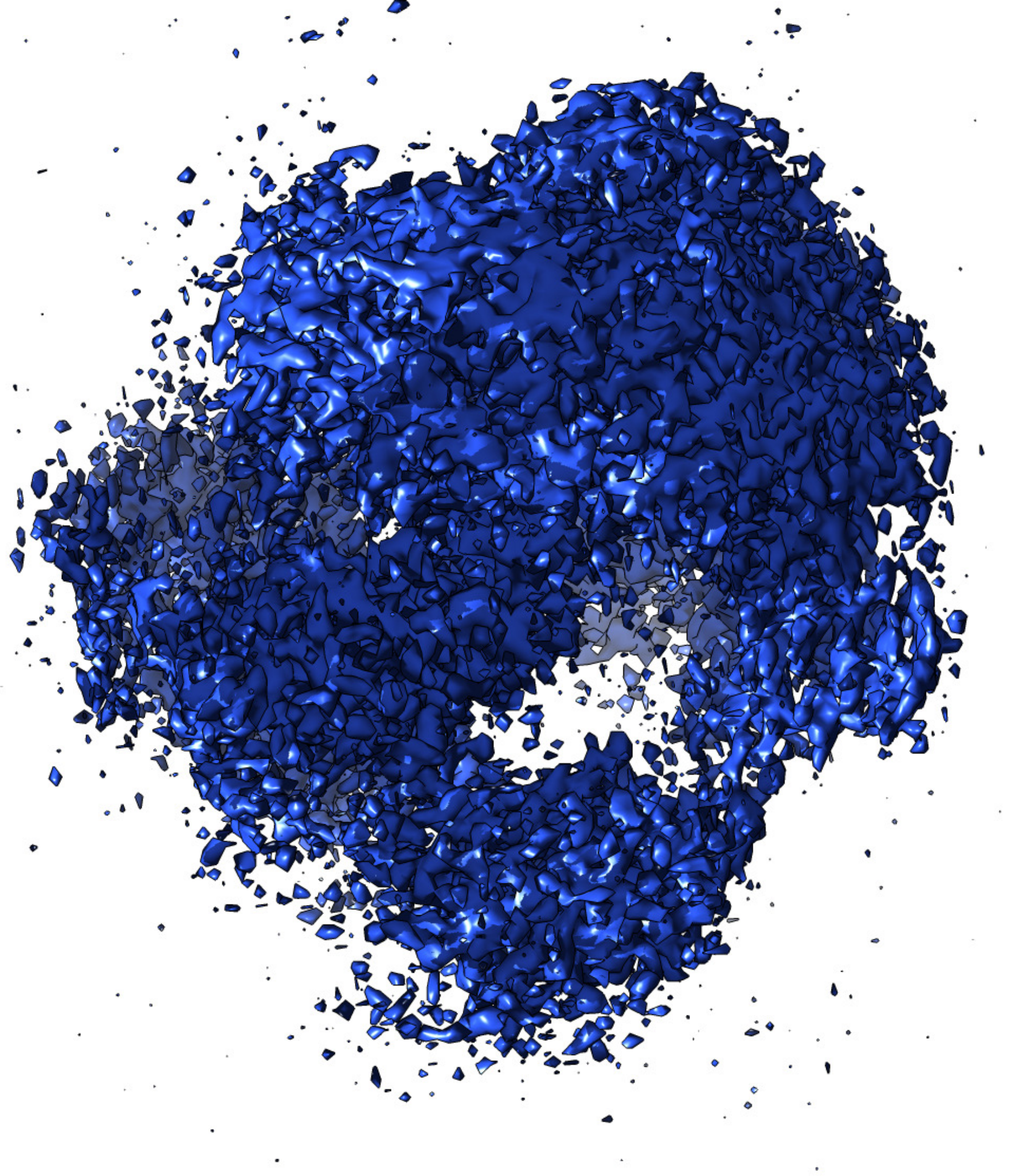}  };
		\node[rotate=270] at (2*\xx+0.,\yy) { \includegraphics[width=\wid]{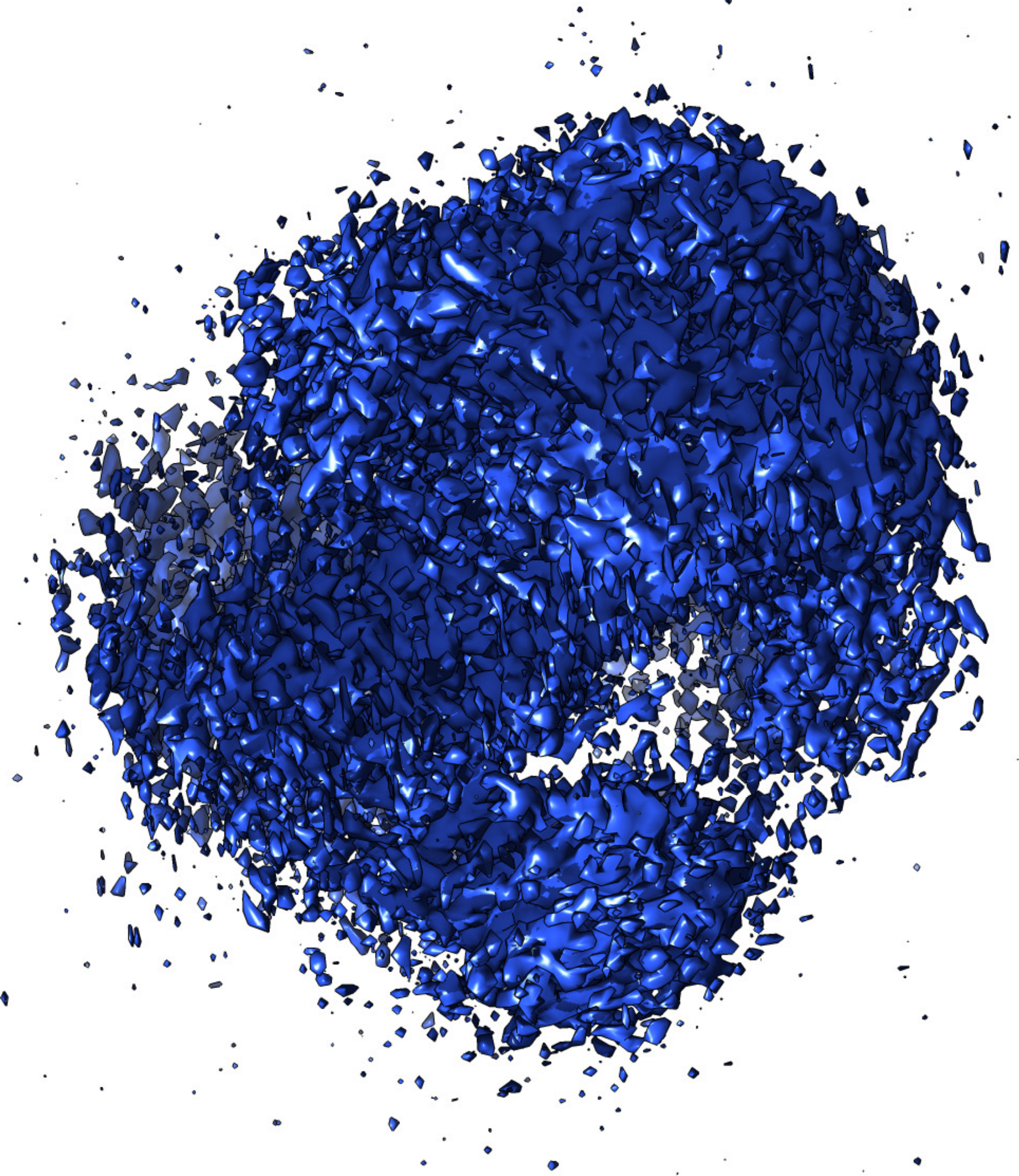}  };

		\spy on (-0.6-1.3*\xx,-0.3+0.5*\yy) in node [left] at (1-1.3*\xx,2+0.5*\yy);
		
		\spy on (-0.6,-0.3) in node [left] at (1,2);
		\spy on (-0.6+1*\xx,-0.3) in node [left] at (1+1*\xx,2);
		\spy on (-0.6+2*\xx,-0.3) in node [left] at (1+2*\xx,2);
		\spy on (-0.6,-0.3+\yy) in node [left] at (1,2+\yy);
		\spy on (-0.6+1*\xx,-0.3+\yy) in node [left] at (1+1*\xx,2+\yy);
		\spy on (-0.6+2*\xx,-0.3+\yy) in node [left] at (1+2*\xx,2+\yy);
		
		\end{tikzpicture} 
	\caption{Visual inspection of the reconstructions at low SNRs. The original volume (gray volume) and the reconstruction using the denoised images by MFVDM (orange volumes) and WP (blue volumes) for three different SNRs: $\{0.3,0.16,0.01\}$. \label{fig:zoom_reconstruction}} 
\end{figure*}

\subsection{WP update}
We train WP on the target graph for 100 epochs using the binary cross entropy loss from \textit{PyTorch} combined with a Sigmoid layer to ensure a score in $[0,1]$.
The target graph is generated by randomly sampling $10,000$ points and adding Gaussian noise with standard deviation $\sigma=0.01$. We found that this small perturbation makes the training more robust to overfitting. We then construct the $K$-nearest neighborhood graph with $K=50$. 
\revision{We train WP with $h=2$. This leads to a subgraph containing about 400 projections. We found that this gives a good trade-off between accuracy and computational time.}  The updated graph is computed using the threshold $\tau=0.5$.

The WP neural network has to be trained only once. We then use it to denoise the graph produced by MFVDM for the three SNRs $\{0.03,0.016,0.01\}$.
The graph rewiring is done by evaluating the WP score for random edges. If the score is below $\tau=0.5$, then the edge is removed. Since the update of one edge changes the WP score of the others, we cannot detect all bad edges at the same time. More precisely, for each node, we select randomly one of its edge and compute its score. Then we iterate until no edges are removed. We display the number of edges removed per iteration in Fig. \ref{fig:updateWP}. We see than after 600 iterations, almost no more edges are removed. 
In the right part of Fig. \ref{fig:updateWP}, we also display the number of edges per node (initially 50). We see that with more noise, more edges are removed on average. We will see in the next paragraphs that this can also be seen as a prior to select a trustworthy subset of nodes.

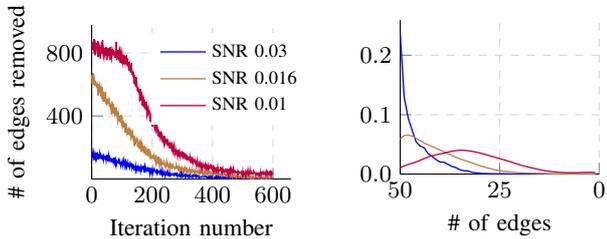
\begin{figure}
    \centering
    \begin{minipage}[t]{0.49\linewidth}
    \centering
    \begin{tikzpicture}     
    \begin{axis}[
          width=1\linewidth, 
          grid=major, 
          grid style={dashed,gray!30}, 
          xlabel= Iteration number,
          ylabel=\# of edges removed,
          xmin = 0,
          ymin = 0,
          ytick={400,800},
          axis x line*=bottom,
          axis y line*=left,
          legend style={at={(1.1,0.95)}, legend cell align=left, align=left, draw=none,font=\scriptsize},
          yticklabel style={
                    /pgf/number format/fixed,
                    /pgf/number format/precision=0,
                    /pgf/number format/fixed zerofill
            },
        scaled y ticks=false
        ]
    \addplot[color=blue,line width=\lw] table [x=x, y=WP, col sep=comma] {Images/XP-15/hist_nb_remove.csv};
    \addlegendentry{SNR 0.03}
    \addplot[color=brown,line width=\lw] table [x=x, y=WP, col sep=comma] {Images/XP-18/hist_nb_remove.csv};
    \addlegendentry{SNR 0.016}
    \addplot[color=purple,line width=\lw] table [x=x, y=WP, col sep=comma] {Images/XP-20/hist_nb_remove.csv};
    \addlegendentry{SNR 0.01}
    \end{axis}
    \end{tikzpicture}
\end{minipage}
\begin{minipage}[t]{0.49\linewidth}
    \centering
    \begin{tikzpicture}     
    \begin{axis}[
          width=1\linewidth, 
          grid=major, 
          grid style={dashed,gray!30}, 
          xlabel= \# of edges,
          xmin = 0, xmax = 50,
          ymin = 0,
          xtick={50,25,0},
          x dir=reverse,
          axis x line*=bottom,
          axis y line*=left,
          legend style={at={(1,0.95)}, legend cell align=left, align=left, draw=none,font=\scriptsize},
          yticklabel style={
                    /pgf/number format/fixed,
                    /pgf/number format/precision=1,
                    /pgf/number format/fixed zerofill
            },
        scaled y ticks=false
        ]
    \addplot[color=blue,line width=\lw] table [x=x, y=WP, col sep=comma] {Images/XP-15/hist_graph_denoising.csv};
    \addplot[color=brown,line width=\lw] table [x=x, y=WP, col sep=comma] {Images/XP-18/hist_graph_denoising.csv};
    \addplot[color=purple,line width=\lw] table [x=x, y=WP, col sep=comma] {Images/XP-20/hist_graph_denoising.csv};
    \end{axis}
    \end{tikzpicture}
    \end{minipage}
    \caption{Properties of the graph rewiring with updates based on WP score of the edges for different SNR.
    Left: number of edges remove in the MFVDM adjacency matrix with respect of the iterations.
    Right: density of the number of edges per node. \label{fig:updateWP}} 
\end{figure}

\subsection{Graph denoising}
Each observation is characterized by its projection axis (see Fig. \ref{fig:cryoEM_pipeline}). The main idea behind constructing a neighborhood graph for denoising is to capture projections that are similar, and thus have similar projection axes. We evaluate the quality of the graph by computing the distance between the viewing directions of every neighbors.
The projection axis of each observation $\yd_m$ can be parameterized by a unitary vector $\xd_m \in \Sc^2$. We then use the geodesic distance on the sphere to evaluate the quality of the neighborhood.
We define the distance between two neighbors $\yd_m$ and $\yd_{m'}$ by
\begin{equation*}
    d_{\mathrm{view}}(\yd_m,\yd_{m'}) = \arccos\left(\langle \xd_m, \xd_{m'} \rangle \right).
\end{equation*}
In Fig. \ref{fig:histogram}, we report the kernel density estimate of the distance between all observations linked by an edge in the MFVDM graph and in the graph updated by WP.
We see that the use of WP produces a neighborhood of projections with more similar viewing directions. This shows that, in the presence of high noise, WP helps to better select edges that are likely to correspond to similar viewing directions.

\begin{figure*}
    \centering
    \begin{tikzpicture}     
    \begin{axis}[
          width=0.3\linewidth, 
          grid=major, 
          grid style={dashed,gray!30}, 
          xlabel= Angle (degrees),
          ylabel=\textbf{SNR:0.03},
          ytick={0,0.04,0.08},
          xmin = 0, xmax = 60,
          ymin=0, ymax=0.099,
          axis x line*=bottom,
          axis y line*=left,
          legend style={at={(1,0.95)}, legend cell align=left, align=left, draw=none,font=\scriptsize},
          yticklabel style={
                    /pgf/number format/fixed,
                    /pgf/number format/precision=2,
                    /pgf/number format/fixed zerofill
            },
        scaled y ticks=false
        ]
    \addplot[color=mycolor1,line width=\lw] table [x=x, y=WP, col sep=comma] {Images/XP-15/hist_out_of_plane.csv};
    \addlegendentry{WP}
    \addplot[color=mycolor2,line width=\lw] table [x=x, y=MFVDM, col sep=comma] {Images/XP-15/hist_out_of_plane.csv};
    \addlegendentry{MFVDM}
    \end{axis}
    \end{tikzpicture}
    \begin{tikzpicture}     
    \begin{axis}[
          width=0.3\linewidth, 
          grid=major, 
          grid style={dashed,gray!30}, 
          xlabel= Angle (degrees),
          ylabel=\textbf{SNR:0.016},
          ytick={0,0.04,0.08},
          xmin = 0, xmax = 60,
          ymin=0, ymax=0.099,
          axis x line*=bottom,
          axis y line*=left,
          legend style={at={(1,0.95)}, legend cell align=left, align=left, draw=none,font=\scriptsize},
          yticklabel style={
                    /pgf/number format/fixed,
                    /pgf/number format/precision=2,
                    /pgf/number format/fixed zerofill
            },
        scaled y ticks=false
        ]
    \addplot[color=mycolor1,line width=\lw] table [x=x, y=WP, col sep=comma] {Images/XP-18/hist_out_of_plane.csv};
    \addplot[color=mycolor2,line width=\lw] table [x=x, y=MFVDM, col sep=comma] {Images/XP-18/hist_out_of_plane.csv};
    \end{axis}
    \end{tikzpicture}
    \begin{tikzpicture}     
    \begin{axis}[
          width=0.3\linewidth, 
          grid=major, 
          grid style={dashed,gray!30}, 
          xlabel= Angle (degrees),
          ylabel=\textbf{SNR:0.01},
          ytick={0,0.04,0.08},
          xmin = 0, xmax = 60,
          ymin=0, ymax=0.099,
          axis x line*=bottom,
          axis y line*=left,
          legend style={at={(1,0.95)}, legend cell align=left, align=left, draw=none,font=\scriptsize},
          yticklabel style={
                    /pgf/number format/fixed,
                    /pgf/number format/precision=2,
                    /pgf/number format/fixed zerofill
            },
        scaled y ticks=false
        ]
    \addplot[color=mycolor1,line width=\lw] table [x=x, y=WP, col sep=comma] {Images/XP-20/hist_out_of_plane.csv};
    \addplot[color=mycolor2,line width=\lw] table [x=x, y=MFVDM, col sep=comma] {Images/XP-20/hist_out_of_plane.csv};
    \end{axis}
    \end{tikzpicture}
    \caption{Estimated density of the angle difference between viewing directions of each pair of nodes connected by an edge.\label{fig:histogram}} 
\end{figure*}
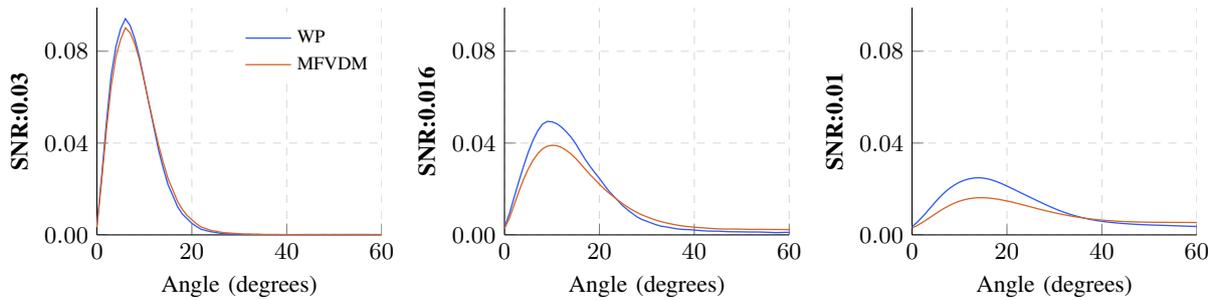

\begin{figure}
    \centering
    \begin{tikzpicture}     
	\node[text width=2cm,text centered,minimum width=2cm,minimum height=2cm] at (1.4,2.7) {high resolution ($n=129$)};
    \begin{axis}[
          width=0.49\linewidth, 
          grid=major, 
          grid style={dashed,gray!30}, 
          xlabel= frequency,
          ylabel=\textbf{SNR: 0.03},
          xtick={10,20},
          xmin = 1, xmax = 20,
          ymin = 0,
          axis x line*=bottom,
          axis y line*=left,
          legend style={at={(0.8,0.45)}, legend cell align=left, align=left, draw=none,font=\scriptsize},
          xticklabel style={
                    /pgf/number format/fixed,
                    /pgf/number format/precision=0,
                    /pgf/number format/fixed zerofill
            },
        scaled x ticks=false
        ]
    \addplot[color=mycolor1,line width=\lw] table [x=x, y=WP, col sep=comma] {Images/XP-15/fsc.csv};
    \addlegendentry{WP}
    \addplot[color=mycolor2,line width=\lw] table [x=x, y=MFVDM, col sep=comma] {Images/XP-15/fsc.csv};
    \addlegendentry{MFVDM}
    \end{axis}   
    \end{tikzpicture}
    \begin{tikzpicture}  
	\node[text width=2cm,text centered,minimum width=2cm,minimum height=2cm] at (1.4,2.7) {low resolution ($n=50$)};   
    \begin{axis}[
          width=0.49\linewidth, 
          grid=major, 
          grid style={dashed,gray!30}, 
          xlabel= frequency,
          xtick={10,20},
          xmin = 1, xmax = 20,
          ymin = 0,
          axis x line*=bottom,
          axis y line*=left,
          legend style={at={(1.2,0.95)}, legend cell align=left, align=left, draw=none,font=\scriptsize},
          xticklabel style={
                    /pgf/number format/fixed,
                    /pgf/number format/precision=0,
                    /pgf/number format/fixed zerofill
            },
        scaled x ticks=false
        ]
    \addplot[color=mycolor1,line width=\lw] table [x=x, y=WP, col sep=comma] {Images/XP-15/fsc_50.csv};
    \addplot[color=mycolor2,line width=\lw] table [x=x, y=MFVDM, col sep=comma] {Images/XP-15/fsc_50.csv};
    \end{axis}   
    \end{tikzpicture}
    \begin{tikzpicture}     
    \begin{axis}[
          width=0.49\linewidth, 
          grid=major, 
          grid style={dashed,gray!30}, 
          xlabel= frequency,
          ylabel=\textbf{SNR: 0.016},
          xtick={10,20},
          xmin = 1, xmax = 20,
          ymin = 0,
          axis x line*=bottom,
          axis y line*=left,
          legend style={at={(1.2,0.95)}, legend cell align=left, align=left, draw=none,font=\scriptsize},
          xticklabel style={
                    /pgf/number format/fixed,
                    /pgf/number format/precision=0,
                    /pgf/number format/fixed zerofill
            },
        scaled x ticks=false
        ]
    \addplot[color=mycolor1,line width=\lw] table [x=x, y=WP, col sep=comma] {Images/XP-18/fsc.csv};
    \addplot[color=mycolor2,line width=\lw] table [x=x, y=MFVDM, col sep=comma] {Images/XP-18/fsc.csv};
    \end{axis}
    \end{tikzpicture}
    \begin{tikzpicture}     
    \begin{axis}[
          width=0.49\linewidth, 
          grid=major, 
          grid style={dashed,gray!30}, 
          xlabel= frequency,
          xtick={10,20},
          xmin = 1, xmax = 20,
          ymin = 0,
          axis x line*=bottom,
          axis y line*=left,
          legend style={at={(1.2,0.95)}, legend cell align=left, align=left, draw=none,font=\scriptsize},
          xticklabel style={
                    /pgf/number format/fixed,
                    /pgf/number format/precision=0,
                    /pgf/number format/fixed zerofill
            },
        scaled x ticks=false
        ]
    \addplot[color=mycolor1,line width=\lw] table [x=x, y=WP, col sep=comma] {Images/XP-18/fsc_50.csv};
    \addplot[color=mycolor2,line width=\lw] table [x=x, y=MFVDM, col sep=comma] {Images/XP-18/fsc_50.csv};
    \end{axis}
    \end{tikzpicture}
    \begin{tikzpicture}     
    \begin{axis}[
          width=0.49\linewidth, 
          grid=major, 
          grid style={dashed,gray!30}, 
          xlabel= frequency,
          ylabel=\textbf{SNR: 0.001},
          xtick={10,20},
          xmin = 1, xmax = 20,
          ymin = 0,
          axis x line*=bottom,
          axis y line*=left,
          legend style={at={(1.2,0.95)}, legend cell align=left, align=left, draw=none,font=\scriptsize},
          xticklabel style={
                    /pgf/number format/fixed,
                    /pgf/number format/precision=0,
                    /pgf/number format/fixed zerofill
            },
        scaled x ticks=false
        ]
    \addplot[color=mycolor1,line width=\lw] table [x=x, y=WP, col sep=comma] {Images/XP-20/fsc.csv};
    \addplot[color=mycolor2,line width=\lw] table [x=x, y=MFVDM, col sep=comma] {Images/XP-20/fsc.csv};
    \end{axis}
    \end{tikzpicture}	
    \begin{tikzpicture}     
    \begin{axis}[
          width=0.49\linewidth, 
          grid=major, 
          grid style={dashed,gray!30}, 
          xlabel= frequency,
          xtick={10,20},
          xmin = 1, xmax = 20,
          ymin = 0,
          axis x line*=bottom,
          axis y line*=left,
          legend style={at={(1.2,0.95)}, legend cell align=left, align=left, draw=none,font=\scriptsize},
          xticklabel style={
                    /pgf/number format/fixed,
                    /pgf/number format/precision=0,
                    /pgf/number format/fixed zerofill
            },
        scaled x ticks=false
        ]
    \addplot[color=mycolor1,line width=\lw] table [x=x, y=WP, col sep=comma] {Images/XP-20/fsc_50.csv};
    \addplot[color=mycolor2,line width=\lw] table [x=x, y=MFVDM, col sep=comma] {Images/XP-20/fsc_50.csv};
    \end{axis}
    \end{tikzpicture}	
    \caption{Fourier shell correlation between the reconstruction and the true volume density. Left: reconstruction with full resolution, $n=129$. Right, reconstruction with low-resolution, $n=50$. \label{fig:fsc}} 
\end{figure}
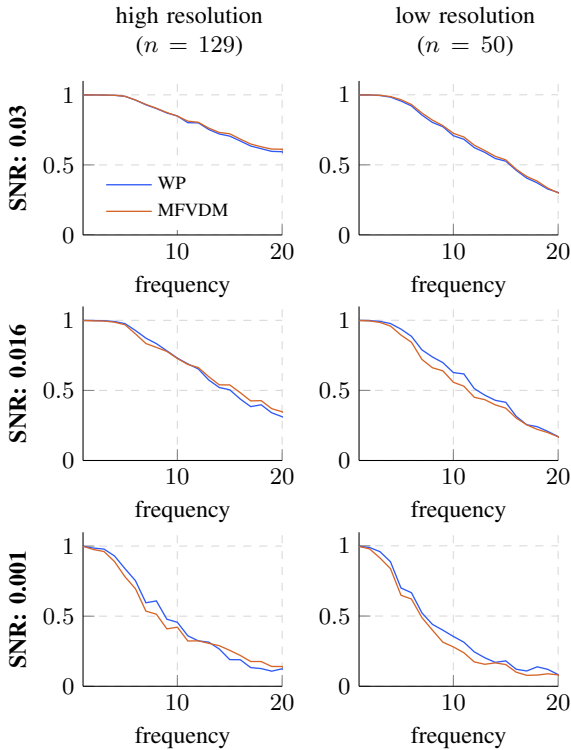

\subsection{Volume reconstruction}

Finally, we show that the improvement in the quality of the graph actually leads to a better reconstruction of the unknown volume.
All reconstructions are performed using the ASPIRE toolbox \cite{zhao2014rotationally} and 1000 denoised images. 
For WP, we make use of the rewiring procedure to select the most trustworthy nodes. We select the 1000 observations with the more remaining edges (at most 50). 

We measure the quality of the reconstruction with the Fourier shell correlation (FSC). The FSC indicates the correlation between the frequencies of the estimated volume and the original. Results are reported in Fig. \ref{fig:fsc}. 

Intuitively, at very low SNR, a better neighborhood graph (see Fig. \ref{fig:histogram}) should translate to improvements at coarse scales. (Note that here we do not perform refinement as would normally be done, since we only want to compare the denoising performance.)
This intuition is confirmed in the FSC plots where WP-denoised graphs yield better scores for lower-resolution volumes (Fig. \ref{fig:fsc}, right column). High frequencies are difficult to recover for both approaches. 

A significant improvement in quality can be observed in the reconstructed 3D volumes in Fig. \ref{fig:zoom_reconstruction}. We see that the shape of the molecule is well retrieved, even at SNR as small as 0.01. Notice also that at SNR 0.01, WP leads to a solid improvement by better retrieving the side structures of the molecule.

\section{Discussion}
We currently use a simple instance of WP to learn a graph prior. On the other hand, one of the strengths of WP is its ability to combine the attributes of each node and the varied connectivity of the graph.
We only use the adjacency matrix to determine the quality of an edge, but we could also leverage node features, for instance by using the inplane angle estimation produced by MFVDM.

Each edge is evaluated by the score function which is trained on the target graph, with a small amount of noise. Unfortunately, when evaluating the score of an edge, the subgraph used probably contains several wrong connections. This introduces a bias in the score function. We believe that training WP on a graph closer to the noisy input used during rewiring can make the evaluation more robust.

Finally, in this paper, we are only removing edges. This makes the image denoising and the reconstruction much simpler to compare with the MFVDM pipeline. However, WP can also be used to get a score on non-existing links, therefore we could detect missing connections. This would require to estimate inplane rotation angle for each new edge. The estimation is likely to be less accurate than the one returned by MFVDM since the latter accounts for a consistency over all the measurements. 

\appendix \section{Proof of Proposition 1}
\subsection{Proof of Proposition \ref{prop:sphere}} \label{app:proof_prop}
\begin{proof}
Let $\thetad\in\SO(3)$ denote a 3D rotation angle, and let $\Rd_\theta\in\R^{3 \times 3}$ denote the corresponding rotation matrix. We follow the caracterization of rotation matrices defined in \cite{sorzano2014interchanging}. Without loss of generality, assume that $\thetad=[\theta_1,\theta_2,\theta_3]$ is an Euler angle given in the ZYZ convention. 

Let $\tilde \xd = [x_1,x_2]\in\R^2$ and $\xd=[\tilde \xd, x_3]\in\R^3$. For any $\thetad=[\theta_1,\theta_2,\theta_3]\in\SO(3)$, we have
\begin{align*}
    y &= \proj \rot(\thetad) \rhod(\xd)\\
      &= \proj \rhod(\Rd^\intercal_\theta\xd)\\
      &= \proj \rhod\left(\Rd_Z^\intercal(\theta_3)\Rd_Y^\intercal(\theta_2) \Rd_Z^\intercal(\theta_1)\xd\right).
\end{align*}
Following \cite{sorzano2014interchanging}, for any $\alpha\in[0,2\pi]$ we have 
\begin{equation*}
    \Rd_Z(\alpha) =  \begin{pmatrix}
                \cos(\alpha) & \sin(\alpha) & 0\\
                -\sin(\alpha) & \cos(\alpha) & 0\\
                0 & 0& 1
              \end{pmatrix}.
\end{equation*}
Let
\begin{equation*}
    \Rd_{2D}(\alpha) =  \begin{pmatrix}
                \cos(\alpha) & \sin(\alpha)\\
                -\sin(\alpha) & \cos(\alpha)
              \end{pmatrix}.
\end{equation*}
Then, for any $\tilde \ud\in\R^2$ and any $\tilde \rhod \in\R^{n\times n \times n}$, we have 
\begin{align*}
    (\proj  \rot([0,0,\theta_3]) \tilde\rhod)(\ud) &=  \proj \tilde\rhod(\Rd_Z(\theta_3) \ud)\\
    &= (\rot_{2D}(\theta_3)\proj \tilde \rhod)(\ud).
\end{align*}
This is the Fourier slice theorem \cite{shkolnisky2012viewing}.

Finally, let $g:[0,\pi]\times [0,2\pi) \to \Sc^2$ denote the mapping such that 
\begin{align*}
 g:(\theta_1,\theta_2)\mapsto \begin{pmatrix} -\cos(\theta_1)\sin(\theta_2)\\ \sin(\theta_1)\sin(\theta_2)\\ \cos(\theta_2) \end{pmatrix}
 \end{align*}
Let $g^{-1}$ denote the inverse mapping, see \cite{sorzano2014interchanging}. Then, setting $\tilde \rhod  = \rot([\theta_1,\theta_2,0])\rhod$ and 
\begin{align*}
    f_\rhod:\quad &  \Sc^2 \to \Lc_2(\R^3)\\
    &\sd \mapsto \proj\rot([g^{-1}\sd, 0]) \rhod),
\end{align*}
lead to 
\begin{equation*}
    y = (\rot_{2D}(\theta_3) f_\rhod( g([\theta_1,\theta_2]) ).
\end{equation*}

Moreover, $\theta=[\theta_{1},\theta_{2},\theta_{3}]\sim\Uc(\SO(3))$ is equivalent to \cite{kuffner2004effective} 
\begin{equation*}
    \theta_1 \sim \Uc([0,2\pi]),\quad \theta_2 \sim \arccos(\Uc([-1,1])), \quad \theta_3 \sim \Uc([0,2\pi]).
\end{equation*}
Standard calculation implies that $g([\theta_1,\theta_2])\sim\Uc(\Sc^2)$.
\end{proof}

\bibliographystyle{apalike}
\bibliography{ref}

\end{document}